\documentclass[aps,prb,twocolumn,amsmath,amssymb,superscriptaddress,floatfix]{revtex4-2}
\usepackage{graphicx}
\usepackage{graphics}
\usepackage{bm}
\usepackage[usenames]{color}
\usepackage{colordvi}
\usepackage{units}
\usepackage{bbm}
\usepackage{booktabs}
\usepackage{array}
\usepackage{placeins}
\usepackage{epsfig}
\usepackage{changes}
\usepackage{braket}
\usepackage{tabularx}
\newcolumntype{L}[1]{>{\raggedright\arraybackslash}p{#1}} 
\newcolumntype{C}[1]{>{\centering\arraybackslash}p{#1}} 
\newcolumntype{R}[1]{>{\raggedleft\arraybackslash}p{#1}} 
\usepackage[colorlinks=true,linkcolor=blue,citecolor=blue,urlcolor=blue]{hyperref}
\newcommand{\be}{\begin{equation}}
\newcommand{\ee}{\end{equation}}
\newcommand{\beqn}{\begin{eqnarray}}
\newcommand{\eeqn}{\end{eqnarray}}

\definecolor{mymagenta}{rgb}{1.0,0.0,1.0}
\definecolor{mycyan}{rgb}{0.0,1.0,1.0}
\definecolor{myyellow}{rgb}{1.0,1.0,0.0}
\definecolor{myorange}{rgb}{1.0,0.27,0.0}

\definecolor{dark-gray}{HTML}{a0a0a0}
\definecolor{dark-red}{HTML}{8b0000}
\definecolor{dark-green}{HTML}{006400}
\definecolor{dark-blue}{HTML}{00008b}
\definecolor{gold}{rgb}{1.0,0.84,0.0}
\definecolor{dark-turquoise}{HTML}{00ced1}

\bibstyle{apsrev.bib}

\begin{document}

\title{Random-bond antiferromagnetic Ising model in a field}
\author{Jean-Christian Angl\`es d'Auriac}
\email{dauriac@neel.cnrs.fr}
\affiliation{Institut N\'eel-MCBT CNRS, B. P. 166, F-38042 Grenoble, France} 
\author{Ferenc Igl{\'o}i}
\email{igloi.ferenc@wigner.hu}
\affiliation{Wigner Research Centre for Physics, Institute for Solid State Physics and Optics, H-1525 Budapest, P.O. Box 49, Hungary}
\affiliation{Institute of Theoretical Physics, Szeged University, H-6720 Szeged, Hungary}
\date{\today}

\begin{abstract}
Using combinatorial optimisation techniques we study the critical properties of the two- and the three-dimensional Ising model with uniformly distributed random antiferromagnetic couplings $(1 \le J_i \le 2)$ in the presence of a homogeneous longitudinal field, $h$, at zero temperature. In finite systems of linear size, $L$, we measure the average correlation function, $C_L(\ell,h)$, when the sites are either on the same sub-lattice, or they belong to different sub-lattices. The phase transition, which is of first-order in the pure system, turns to mixed order in two dimensions with critical exponents $1/\nu \approx 0.5$ and $\eta \approx 0.7$. In three dimensions we obtain $1/\nu \approx 0.7$, which is compatible with the value of the random-field Ising model, but we cannot discriminate between second-order and mixed-order transitions.
\end{abstract}


\maketitle

\section{Introduction}
\label{sec:intro}

Phase transitions in systems with quenched disorder are not well understood despite intensive research. Exact results in this field are very scarce. In experiments thermal equilibrium is difficult to reach, which is also true for several numerical simulations. A paradigmatic system in this field of research is the random-field Ising model (RFIM)\cite{nattermann,belanger}, for which a phase-transition takes place in three dimensions ($d=3$), while in $d=2$ the random field destroys the transition which takes place in the pure model. The phase transition in the $d=3$ RFIM is governed by a zero temperature fixed point, the properties of which can be very efficiently studied by combinatorial optimisation algorithms\cite{ogielski,angles,middleton,fytas}. In this way the ground states of the random samples can be exactly calculated and the simulation is performed for large systems with high statistics.

In a theoretical point of view the perturbative renormalization group (PRG) can be carried in all orders of perturbation theory for the RFIM\cite{parisi1,parisi2}. It predicts dimensional reduction, which means that the critical exponents of the RFIM in $d$ dimensions are the same as the exponents of the pure Ising model in $d-2$ dimensions. Another prediction of the PRG is that the RFIM and the disordered antiferromagnetic Ising model in an external magnetic field (DAFF) are in the same universality class, which means that critical exponents and other critical parameters are the same for the two models and they do not depend on the specific form of disorder. This statement is first formulated for random bonds\cite{fishman}, but it has afterwards been generalised for dilution\cite{cardy}. The diluted version is very important, since it can be connected with experiments, which has been performed extensively\cite{belanger}. Regarding the predictions of the PRG, some are false c.f. the dimensional reduction\cite{bricmont}, but some turned to be true, regarding universality of the RFIM and the diluted antiferromagnetic Ising model in a field.

In this paper we are going to study the critical properties of the antiferromagnetic Ising model in an external magnetic field with random couplings at zero temperature in $d=2$ (square lattice) and in $d=3$ (simple cubic lattice). In the pure model in the ground state there is a first-order transition and at the transition point the ground state is infinitely degenerate. Switching on disorder the degeneracy at the transition point is lifted and the properties of the transition are expected to be changed. For the numerical calculations we use very efficient combinatorial optimisation algorithms and calculate the exact ground state of large finite samples. We are going to obtain precise numerical estimates for the critical properties of the $d=2$ model, less extensive simulations will be performed in $d=3$. Regarding previous studies: a very large number of numerical simulations have been performed for the RFIM, only a few simulations are devoted to the diluted antiferromagnetic Ising model in a field\cite{sourlas,hartmann,ahrens,fernandez,picco} and we are not aware of simulations for the random-bond antiferromagnetic Ising model in a field. In this paper we aim to fill this gap.

The rest of the paper is organised in the following way. The model and the method of investigation is presented in Sec.\ref{sec:model}. The numerical results are calculated and analysed in Sec.\ref{sec:results} and a discussion is presented in Sec.\ref{sec:disc}.

\section{The model}
\label{sec:model}

Our starting point is the antiferromagnetic Ising model in a field:
 
\begin{align}
\begin{split}
{\cal H}_{\rm AF}=J\sum_{\langle i,j \rangle} \sigma_{i} \sigma_{j}-h\sum_{i} \sigma_{i}\;.
\label{Hamilton}
\end{split}
\end{align}
in terms of $\sigma_i=\pm 1$ at site $i$ and the first sum runs over nearest neighbours. We
consider finite lattices of linear size $L$ and with periodic boundary conditions (PBC-s). The ground state of this pure model is antiferromagnetic for $h<dJ$ and ferromagnetic for $h>dJ$. The transition at $h_c=dJ$ is of first order and at this point the ground state is infinitely degenerate: each spins in one of the sub-lattices can be either $\sigma_i=1$ or $\sigma_i=-1$.

For computational reasons it is convenient to perform a gauge transformation $\sigma_i \to (-1)^{\sum_k i_k}\sigma_i$, with $i=\{i_1,i_2,\dots,i_d\}$. For a hypercubic lattice with PBC and even $L$ we obtain a ferromagnet in an alternating magnetic field:
\begin{align}
\begin{split}
{\cal H}_{\rm AF}=-J\sum_{\langle i,j \rangle} \sigma_{i} \sigma_{j}-h\sum_{i} (-1)^{\sum_k i_k} \sigma_{i}\;.
\label{Hamilton_F}
\end{split}
\end{align}
Next step we consider disorder in the system. In the literature one generally considers random site dilution, which amounts to replace the spin variable $\sigma_i$ by $\sigma_i\varepsilon_i$, where the $\varepsilon_i$ are independent random numbers: $\varepsilon_i=0$ ($\varepsilon_i=1$), with probability $p$ ($1-p$). Instead we consider here the model with random couplings:
\begin{align}
\begin{split}
{\cal H}_{\rm RAFF}=-\sum_{\langle i,j \rangle} J_{ij} \sigma_{i} \sigma_{j}-h\sum_{i} (-1)^{\sum_k i_k} \sigma_{i}\;.
\label{Hamilton_RAFF}
\end{split}
\end{align}
and the $J_{ij}$ couplings are taken independently from the box-like distribution:
\begin{align}
  \begin{split}
    \pi(J) &=
    \begin{cases}
      (\Delta J)^{-1} & \hspace*{0.65cm}\text{for } J_0<J\le J_0+\Delta J\;,\\
      0 & \hspace*{0.65cm}\text{otherwise.}
    \end{cases}
  \end{split} 
  \label{eq:J_distrib} 
\end{align}
In the limit $\Delta J \to 0$ we have the pure system.
In the following we argue, that the phase-diagram of the random system can be different, if the smallest coupling is $J_{min}=J_0>0$ or $J_{min}=J_0=0$. Indeed, the ground state is strictly AFM ordered, if $h<2d J_{min}$ and for $h>2d J_{min}$ excitations destroy (locally) the AFM order. In our numerical work we shall investigate the region with $J_0>0$ and keep $J_0=\Delta J=1$. 
\begin{figure}[h!]
\begin{center}
\includegraphics[width=1.\linewidth]{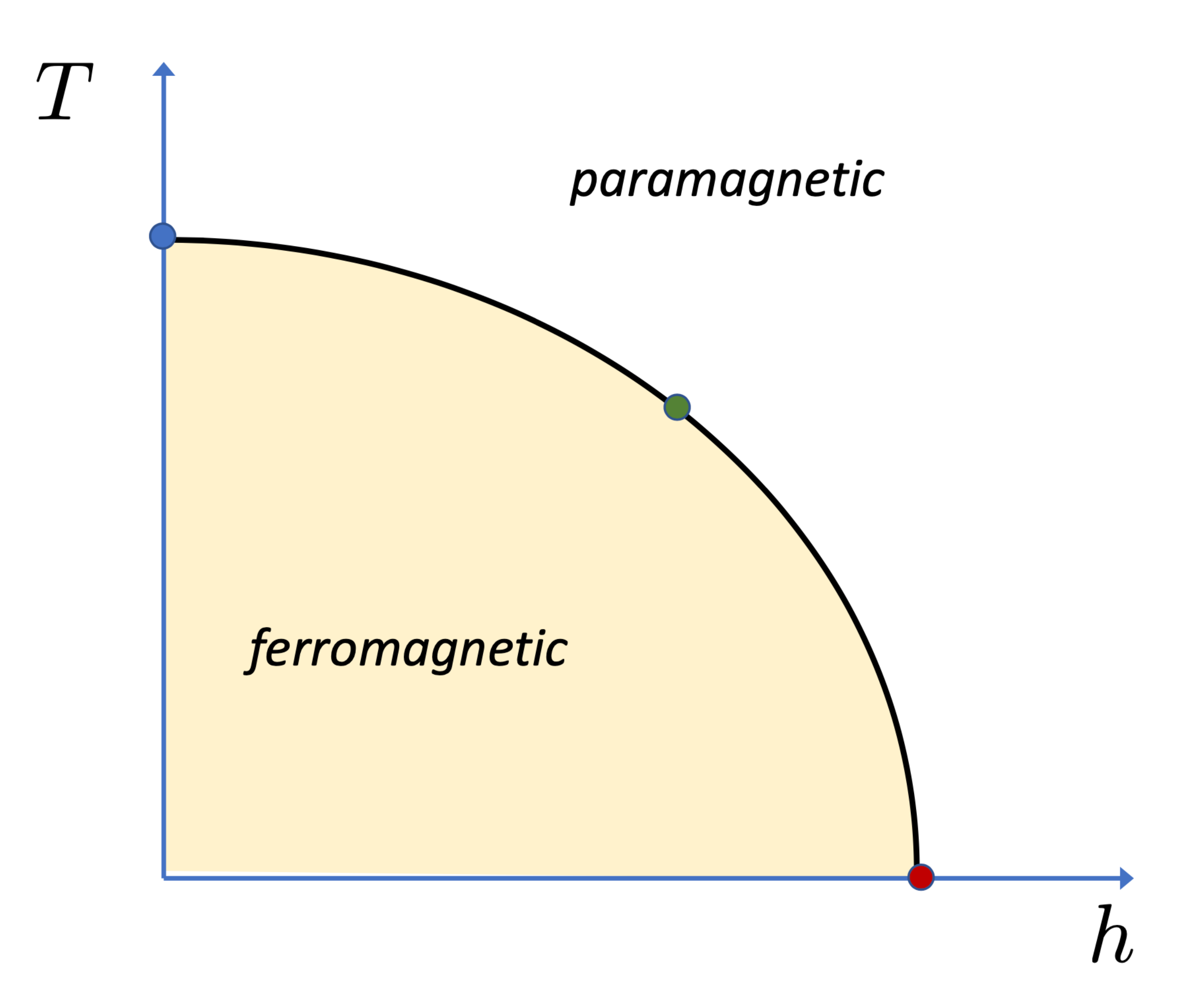}
\end{center}
\caption{\label{fig:phase}(Color online) Schematic phase-diagram of the RAFF in $d=3$. In the zero field fixed point (blue) the critical exponents are those of the random-bond Ising model. For finite value of the field the transition is expected to be controlled by the fixed point of the random-field Ising model (green) according to the prediction of the perturbative renormalization group. In $d=2$ the ferromagnetic phase for $T>0,h>0$ is expected to be absent. The properties of the model at the zero temperature (red) fixed point are the subject of this paper.}
\end{figure}

In the plane of temperature ($T$) and field ($h$) the schematic phase-diagram of the RAFF is shown in Fig.\ref{fig:phase}, which in $d=3$ contains a ferromagnetic and a paramagnetic phase. With zero field, $h=0$, the phase diagram has a random-bond Ising fixed point, in which the critical exponents\cite{hasenbusch} are different from their values in the pure model in $d=3$. For finite field, according to PRG arguments, the transition is controlled by the fixed point of the random-field Ising model. Finally, there is a zero-temperature fixed point, the properties of which will be studied in this paper. In $d=2$ the critical exponents at the zero-field fixed point are the same as for the pure Ising model, however with logarithmic corrections\cite{dotsenko}. In $d=2$ the ferromagnetic phase for $T>0,h>0$ is expected to be absent, which follows from the PRG results. In this paper, we focus on calculating the critical properties of the RAFF at the zero-temperature fixed point.

\section{Results at zero temperature}
\label{sec:results}

At zero temperature the ground state of finite samples has been calculated exactly by a very efficient combinatorial optimisation algorithm. The problem is turned into the so-called max-flow problem \cite{angles85}: all the sites of the first sub-lattice are linked to an extra site (the source) by a bond weighted by $h$, while the sites of the other sub-lattice are linked to another extra site (the sink) also weighted by $h$. The bond between two original sites $i$ and $j$ are weighted by $J_{ij}$. Then the min cut separating the source from the sink realizes one of the possibly many ground states. These min cut is found via the max flow algorithm using the Goldberg and Tarjan algorithm \cite{goldberg}.

We calculated average correlation functions:
\be
C_L(\ell)=\overline{\langle \sigma_i \sigma_{i+\ell}} \rangle\;,
\label{C}
\ee
with $\ell=\{\ell_1,\ell_2,\dots,\ell_d\}$ and
$\langle \dots \rangle$ denotes average in the ground state of a given sample, which amounts to averaging for all spin-pairs having a distance $|\ell|$, and $\overline{\cdots}$ stands for the average over quenched disorder. We concentrate on the behavior of $C_L(\ell)$ for the largest separations and calculate $C_L(\ell_{max}^{(\uparrow \downarrow)}) \equiv C^{(\uparrow \downarrow)}_L$, when the sites are on different sub-lattices, $\ell_{max}^{(\uparrow \downarrow)}=\{L/2-1,L/2,\dots,L/2\}$ and $C_L(\ell_{max}^{(\uparrow \uparrow)}) \equiv C^{(\uparrow \uparrow)}_L$, when the sites belong to the same sub-lattice, $\ell_{max}^{(\uparrow \uparrow)}=\{L/2,L/2,\dots,L/2\}$.

\subsection{Square lattice}

\begin{figure}[h!]
\begin{center}
\includegraphics[width=1.\linewidth]{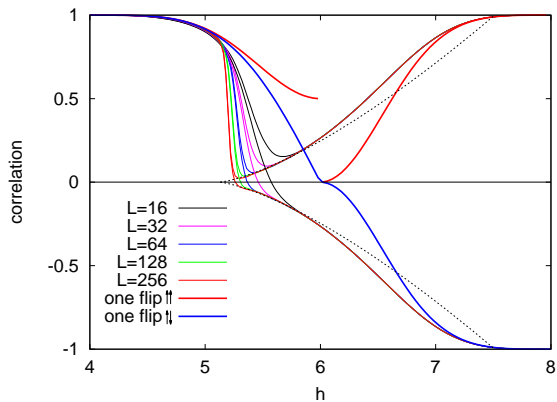}
\end{center}
\caption{\label{fig:corr_square}(Color online) Average correlation functions, $C_L^{(\uparrow \uparrow)}(h)$  (upper curves) and $C_L^{(\uparrow \downarrow)}(h)$ (lower curves) as a function of the field, $h$, calculated on finite square lattices. Results of the one spin-flip approximation are shown by full lines. The dashed lines illustrate the expected limiting behaviour in the vicinity of the transition point in the thermodynamic limit, see in Eq.(\ref{beta}).}
\end{figure}

\begin{figure}[h!]
\begin{center}
\includegraphics[width=1.\linewidth]{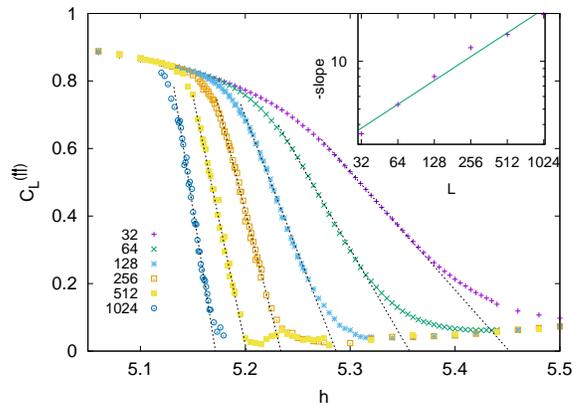}
\end{center}
\caption{\label{fig:C1}(Color online) $C_L^{(\uparrow \uparrow)}(h)$ at the decreasing parts and their slope at the inflexion points are indicated by straight dashed lines. The crossing point of a straight line with the horizontal axis defines the finite-size transition parameter $h_0(L)$. From right to left $L=32,64,128,256,512$ and $1024$.  In the inset the slopes of the curves at the inflexion point are plotted versus $L$ in a double logarithmic scale. The slope of the dashed straight line is $\epsilon=1/2$.}
\end{figure}

For the square lattice we considered finite lattices of linear size: $L=16,32,64,128,256,512$ and $1024$ and the number of realizations varied between $10000$, for the smaller sizes and $500$, for the larger ones. We have calculated the average correlation functions $C^{(\uparrow \downarrow)}_L(h)$ and $C^{(\uparrow \downarrow)}_L(h)$, which are shown in Fig.\ref{fig:corr_square}.
%
For relatively smaller values of $h <5$, the numerical curves are very close to one another and their values practically agree with those calculated within the one spin-flip approximation. Having a closer look to the curves one can see, that in this region for a fixed $h$, $C^{(\uparrow \uparrow)}_L(h)$ and $C^{(\uparrow \downarrow)}_L(h)$ monotonously increase with $L$. If the value of $h$ is increased further, the curves for different lengths start to cross each other, e.g. $C^{(\uparrow \downarrow)}_{2L}(h)=C^{(\uparrow \downarrow)}_L(h)$ at $h=h^*(L)$. Generally $h^*(L)$ decreases with increasing $L$, but converge rapidly to a limiting value: $\lim_{L \to \infty} h^*(L)=h^*$. This limiting value of $h^*$ looks identical for $C^{(\uparrow \downarrow)}_L(h)$, too. After passing the crossing points the order of the the curves for different values of $L$ reverses, and their values start to decrease rapidly and exceed a minimum. The value at the minimum tends towards zero in the large $L$ limit. In Fig.\ref{fig:C1} we enlarge the sloping part of the curves. As can be seen in the figure, the finite size curves run over an inflection point at $h=\tilde{h}(L)$, at which point we draw a tangential straight line described by the equation: $y=C' \times (h-h_0)$. Here
\be
C'=C'(L)=\left.\frac{{\rm d} C}{{\rm d} h}\right|_{\tilde{h}(L)}\;
\ee
is the slope and $h_0=h_0(L)$ is the crossing point with the horizontal axis, which can be used as a finite-size transition parameter. By inspection we notice a power-law variation: $C'(L) \sim L^{\epsilon}$, with $\epsilon \approx 0.5$, see in the inset of Fig.\ref{fig:C1}.
%

\begin{figure}[h!]
\begin{center}
\includegraphics[width=1.\linewidth]{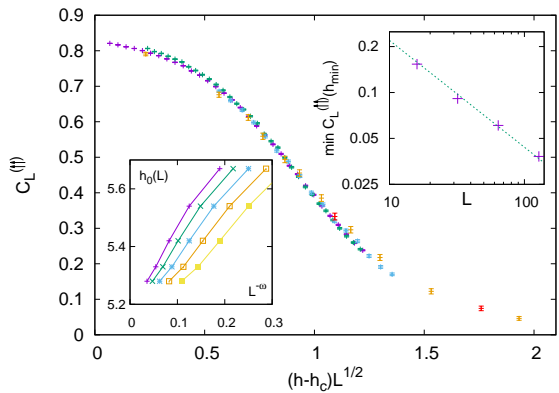}
\end{center}
\caption{\label{fig:C_scale1}(Color online) The average correlation function, $C_L^{(\uparrow \uparrow)}(h)$, close to the transition point as the function of the scaling variable $(h-h_c)L^{1/2}$, with $h_c=5.102$ estimated in the inset. ($L=128$ \textcolor{magenta}{+}, $L=256$ \textcolor{green}{+}, $L=512$ \textcolor{blue}{+}, $L=1024$ \textcolor{yellow}{+}, $L=2056$ \textcolor{red}{+}.)In left the inset the finite-size transition parameter $h_0(L)$, as defined in Fig.\ref{fig:C1} is plotted as function of $L^{-\omega}$, for different values of $\omega=0.6,0.55,0.5,0.45$ and $0.4$, from top to bottom. The best asymptotic form is obtained with $\omega \approx 0.5$, leading to an estimate for $h_c$, used in the main panel. Right inset: finite-size scaling of the minimum value of $C^{(\uparrow \uparrow)}_L(h_{min})$ in log-log plot. The slope of the dashed straight line is $2x \approx 0.7$.}
\end{figure}

Consequently in the thermodynamical limit the slope of the curve diverges and at the same time the extension of the critical region, $\Delta h(L) \sim h_0(L) - h_c$, with $h_c=\lim_{L \to \infty} h_0(L)$ being the transition point, shrinks to zero. According to the inset of Fig.\ref{fig:C_scale1} this relation is also in a power-low form: $\Delta h(L) \sim L^{-\omega}$, with $\omega \approx 0.5$. Using the corresponding scaling combination, $(h-h_c)L^{0.5}$, the finite-size correlation functions can be put approximately to a master curve, as shown in the main panel of Fig.\ref{fig:C_scale1}. 

%

In a finite system at the critical point the correlation length, $\xi$, is limited by the linear size of the system, $\xi \sim L$, and the extension of the critical region scales as $\Delta(h) \sim \xi^{-1/\nu}$. Hence the correlation length critical exponent in our case is $\nu=1/\omega \approx 2$. In the thermodynamical limit the two limiting transition points become equal: $h^*=h_c$ and at the transition point $C^{(\uparrow \uparrow)}_L(h)$, as well as $C^{(\uparrow \downarrow)}_L(h)$ has a jump, from a finite value at $h \to h^*$ to zero at $h \to h_c$. At the right side of the transition point for $h>h_c$ the transition is continuous, which is manifested by the fact that the value of the minima of $C^{(\uparrow \uparrow)}_L(h)$ goes to zero as a power-law: $C^{(\uparrow \uparrow)}_L(h_{min}) \sim L^{-2x}$. This is checked in the right inset of Fig.\ref{fig:C_scale1} and an estimate $2x \approx 0.7$ is obtained.

If we consider the behaviour of the correlation functions in the thermodynamic limit, then for $h<h_c$ we have $C^{(\uparrow \uparrow)}(h)=C^{(\uparrow \downarrow)}(h)$ and at $h=h_c$ there is a finite jump to zero. At the other side of the transition point $h>h_c$ we have $C^{(\uparrow \uparrow)}(h)=-C^{(\uparrow \downarrow)}(h)$ and close to the transition point the variation is of a power-law form:
\be
C^{(\uparrow \uparrow)}(h) \sim (h-h_c)^{2\beta},\quad h>h_c \;,
\label{beta}
\ee
with $\beta=x \nu$. With our previous estimates we have $2\beta \approx 1.4$ and we illustrate such type of a behavior in Fig.\ref{fig:corr_square}.

We can thus conclude that the transition of the RAFF in $d=2$ and in zero temperature is of mixed order. Mixed-order transitions can be observed in a variety of models\cite{mukamel}, also the RAFF in $d=1$ has a mixed-order transition\cite{lajko}.

\subsection{Cubic lattice}
\label{sec:cubic}

\begin{figure}[h!]
\begin{center}
\includegraphics[width=1.\linewidth]{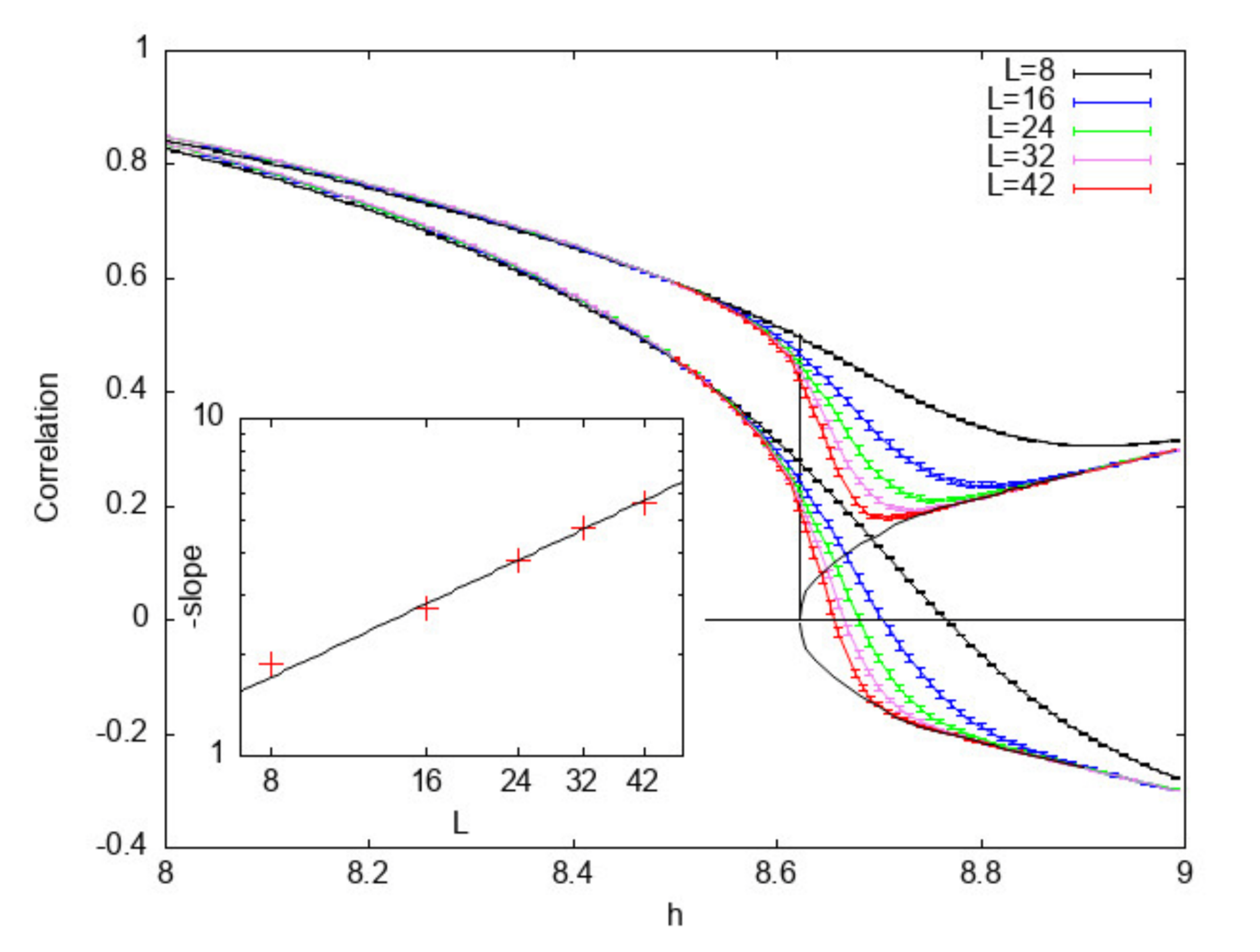}
\end{center}
\caption{\label{fig:corr3dd}(Color online) Average correlation functions $C^{(\uparrow \uparrow)}_L(h)$ (upper curves) and  $C^{(\uparrow \downarrow)}_L(h)$ (lower curves) for different sizes for the cubic lattice. If mixed-order transition takes place the dashed lines indicate the possible behaviour of the curves in the thermodynamic limit. Inset: slope of $C^{(\uparrow \downarrow)}_L(h)$ at the crossing point with the $x$-axis as a function of the linear size of the cubic lattice, $L$, in a log-log plot. The slope of the dashed line is $\varepsilon \approx 0.7$.}
\end{figure}

For the cubic lattice we used finite systems of linear size $L=8,16,24,32$ and $42$ with periodic boundary condition and the number of samples was at least $1000$ even for the largest systems. We calculated the average correlation functions, $C^{(\uparrow \uparrow)}_L(h)$ and $C^{(\uparrow \downarrow)}_L(h)$ which are shown in Fig.\ref{fig:corr3dd} for different sizes.
%
Comparing the position of the curves it is seen that (for a given $L$) $C^{(\uparrow \downarrow)}_L(h)$ is considerably shifted down from $C^{(\uparrow \uparrow)}_L(h)$. This is different from the $d=2$ case, when the limiting positions of $C^{(\uparrow \uparrow)}_L(h)$ and $C^{(\uparrow \downarrow)}_L(h)$ in the thermodynamic limit looks identical, see in Fig.\ref{fig:corr_square}.
Considering the relative positions of the finite-size curves, (separably for $C^{(\uparrow \downarrow)}_L(h)$ and $C^{(\uparrow \downarrow)}_L(h)$), these are similar to the $d=2$ case. For $h<8.5$ these are very close to each other, but at the same time the correlations are larger for larger $L$-s. Increasing the value of $h$ the correlation curves for different sizes cross each other, and these crossing points seem to approach a limiting value $h^*$ for large values of $L$. This limiting crossing point seems to be the same for both $C^{(\uparrow \uparrow)}_L(h)$ and $C^{(\uparrow \downarrow)}_L(h)$. Passing the crossing point, for $h>h^*$ the curves start to decrease rapidly in an approximately linear form and the slope of these linear parts increases with $L$. By inspection the slopes of the curves for a given $L$ are close to each other for $C^{(\uparrow \uparrow)}_L(h)$ and for $C^{(\uparrow \downarrow)}_L(h)$.

In the following let us concentrate on the fast varying behaviour of $C^{(\uparrow \downarrow)}_L(h)$ and let us define a finite-size transition parameter with the position of the crossing point of the curve with the $x$-axis, which is denoted by $h_0(L)$. The slope of the curves at $h_0(L)$ is denoted by $C'(L)$, which are plotted as a function of $L$ in Fig.\ref{fig:slope1} in a log-log scale.

According to this figure the slopes have a power-law size-dependence: $C'(L) \sim L^{\varepsilon}$, with $\varepsilon \approx 0.7$. Assuming that this behaviour remains valid even at the thermodynamic limit, then the slope at a true transition point, defined as $h_c=\lim_{L \to \infty} h_0(L)$, will be infinite. Next we have to decide about the behaviour of the correlations at the transition point. One possibility, that there is a finite jump, like in $d=2$, and the transition is of mixed order. In this case we have the relation: $1/\nu=\varepsilon \approx 0.7$. In this case for $h>h_c$ the correlation functions are expected to follow a power-law dependence, like in Eq.(\ref{beta}). Since the jump in finite systems is relatively small, one can expect that this jump vanishes in the thermodynamic limit and the transition is of second order, however with a very small value of the magnetisation exponent, $\beta$. This scenario would fit to the prediction of the PRG, that the critical properties of the RAFF are the same as that of the RFIM, for the latter the critical exponents being $1/\nu \approx 0.7$ and $\beta \approx 0$. With our limited finite-size date we cannot discriminate between these two possibilities for the type of the phase transition in the system.

\section{Discussion}
\label{sec:disc}
In this paper we considered the antiferromagnetic Ising model with random couplings and in the presence of a homogeneous field and studied the properties of the phase transition at zero temperature. Using very efficient numerical algorithms we calculated exact ground states of finite hypercubic lattices in $d=2$ and $d=3$ for a large set of random samples. We have calculated average spin-spin correlation functions and studied their properties, when the two sites are at the same sub-lattice or the sites belong to different sub-lattices. The phase transition in the non-random system is of first order and the ground state at the transition point being infinitely degenerate with a finite entropy per site. Due to disorder, this degeneracy is lifted and the transition turns to mixed-order in $d=2$. The critical exponents are $1/\nu \approx 0.5$ and $\eta=2x \approx 0.7$, which represents a new random universality class. In $d=3$ we could not decide between mixed-order or second-order transition due to our limited finite-size results. The second-order scenario would fit to the prediction of the PRG theory, having critical exponents $1/\nu \approx 0.5$ and $\beta \approx 0$.

An interesting question, what happens with the transition at finite temperature. According to the prediction of the PRG the critical properties of the RAFF should be the same as that of the RFIM. In this way in $d=2$ there should be no ordered phase, while in $d=3$ the critical exponents should be the same as for the RFIM, both at $T=0$ and for $T>0$. For the diluted Ising antiferromagnet in a field this scenario has been numerically confirmed in $d=3$\cite{picco}, while the case $d=2$ has not yet been studied.

Another way to complete our model is to introduce (random) transverse fields into the problem. This question has been studied in $d=1$ and reentrant critical behaviour is observed around the RAFF fixed point\cite{lajko}. Similar type of reentrant phase transitions are expected to take place in higher dimensions, too.

\section*{Appendix: Correlations in the one spin-flip approximation}

Here we consider the square lattice, the considerations are straightforward to generalise for the cubic lattice.
Let us consider the ferromagnetic phase with $h < 4$ and start to increase the staggered longitudinal field over $h=4$. Here we consider such processes, when single $\uparrow$ spins are flipped in one of the sub-lattices. The ferromagnetic ground state will then change locally at such a site, where the strength of the field exceeds the sum of the four local couplings:
\be
\sum_{j=1}^4 J_{ij}<h\;.
\ee
At this point the originally $\uparrow$ spin will turn to $\downarrow$.
The probability distribution, $P_4(x) \rm{d}x$ of the sum of four random couplings: $x=J_1+J_2+J_3+J_4$ is given by the convolution:
\begin{align}
P_4(x)&=\int \rm{d} J_1 \int \rm{d} J_2 \int \rm{d} J_3 \pi_1(J_1)\pi_1(J_2)\pi_1(J_3) \nonumber \\
&\times \pi_1(x-J_1-J_2-J_3)\;,
\end{align}
and can be calculated by the box-distribution in Eq.(\ref{eq:J_distrib}) (with $J_0=\Delta J=1$):
\begin{align}
P_4(x)&=\frac{1}{12}\left[|8-x|^3-4|7-x|^3+6|6-x|^3 \right. \nonumber \\
&-\left. 4|5-x|^3+|4-x|^3\right]\;.
\label{P_4(x)}
\end{align}
This is non-zero for $4<x<8$, and in the range $4<x<6$ is given by:
\be
P_4(x)=
\begin{cases}
\dfrac{1}{6}(x-4)^3,\quad &4<x<5\\
\\
\dfrac{1}{6}(x-4)^3-\dfrac{2}{3}(x-5)^3,\quad &5<x<6\;
\end{cases}
\ee
and symmetric for $x=6$.

The integrated density, $\mu_4(x)$ in this range behaves as:
\begin{align}
\mu_4(x)&=\int_4^x P_4(x'){\rm d}x' \nonumber \\
&=
\begin{cases}
\dfrac{1}{24}(x-4)^4,\quad 4<x<5\\
\\
\dfrac{1}{24}(x-4)^4-\dfrac{1}{6}(x-5)^4,\quad 5<x<6\;.
\end{cases}
\end{align}
The average number of flipped $\uparrow$ spins at a field $h$ is given by:
\be
n_{fl}=\mu_4(h)\frac{L^2}{2}=\frac{1}{48}(h-4)^4 L^2,\quad 4<h<5\;,
\ee
and the average correlation function between two sites having a distance $L/2-1$:
\begin{align}
\label{C(L/2-1)}
&C_L^{(\uparrow\uparrow)}(h)=\overline{\langle \sigma_1^z\sigma_{L/2}^z \rangle}=\frac{L^2/2-2n_{fl}}{L^2/2} \nonumber \\
&=
\begin{cases}
1-\dfrac{1}{12}(h-4)^4,\quad 4<h<5\\
\\
1-\dfrac{1}{12}(h-4)^4+\dfrac{1}{3}(h-5)^4,\quad 5<h<6\;.
\end{cases}
\end{align}
It has the symmetry: $C_L^{(\uparrow\uparrow)}(h)=-C_L^{(\uparrow\uparrow)}(12-h)$.

The average correlation function between two sites at a distance $L/2$ is given by:
\begin{align}
\label{C(L/2)}
&C_L^{(\uparrow\downarrow)}(h)=\overline{\langle \sigma_{L/2}^z\sigma_L^z \rangle}=\frac{L^2/2+\left(1-2\mu_4(h)\right)^2L^2/2}{L^2} \nonumber \\
&=
\begin{cases}
\dfrac{1}{2}\left[1+\left(1-\dfrac{1}{12}(h-4)^4\right)^2\right],\quad 4<h<5\\
\\
\dfrac{1}{2}\left[1+\left(1-\dfrac{1}{12}(h-4)^4+\dfrac{1}{3}(h-5)^4\right)^2\right],\quad 5<h<6\;.
\end{cases}
\end{align}

\end{document}